\documentclass[12pt]{article}
\usepackage{epsfig}
\usepackage{float}
\usepackage{graphicx}
\begin{document}
\rightline{NKU-2016-SF6}
\bigskip

\newcommand{\be}{\begin{equation}}
\newcommand{\ee}{\end{equation}}
\newcommand{\noi}{\noindent}
\newcommand{\ra}{\rightarrow}
\newcommand{\bib}{\bibitem}
\newcommand{\refb}[1]{(\ref{#1})}
\newcommand{\bff}{\begin{figure}}
\newcommand{\eff}{\end{figure}}
\newcommand{\bigd}{\bigtriangledown}

\begin{center}
{\Large\bf Electromagnetic  perturbations of a de Sitter \\
black hole in massive gravity}

\end{center}
\hspace{0.4cm}
\begin{center}
Sharmanthie Fernando \footnote{fernando@nku.edu} \& Amanda Manning\footnote{manninga7@mymail.nku.edu}\\
{\small\it Department of Physics, Geology \& Engineering Technology}\\
{\small\it Northern Kentucky University}\\
{\small\it Highland Heights}\\
{\small\it Kentucky 41099}\\
{\small\it U.S.A.}\\

\end{center}

\begin{center}
{\bf Abstract}
\end{center}
The main purpose of this paper is to study quasinormal modes (QNM)  of a  black hole in massive gravity with a positive cosmological constant due to electromagnetic  perturbations. A detailed study of the QNM frequencies for the 
electromagnetic field is done by  varying the parameters of the theory such as the mass, scalar charge, cosmological constant, and the spherical harmonic index. We have employed the sixth order WKB approximation to calculate the QNM frequencies.  The electromagnetic potential for the near extreme massive gravity de Sitter black hole is approximated with the  P$\ddot{o}$schl-Teller potential to obtain exact frequencies.  The null   geodesics of the black hole in massive gravity  is employed to describe the absorption cross sections  at high-frequency limit.

\hspace{0.7cm}


\section{ Introduction}

The universe is expanding at an accelerated rate as is supported by many observations  \cite{perl}\cite{riess}\cite{sper}\cite{teg}\cite{sel}. The concept of dark energy is introduced to justify this accelerated expansion. The simplest model for dark energy is the well known cosmological constant  with the equation of state given by $\omega =-1$. This implies that the universe we live is asymptotically de Sitter. Hence understanding physics in a de Sitter geometry becomes all the more important. From the point of string theory which is the most popular approach to quantum gravity, there are many advancements in anti-Sitter space unlike the de Sitter space. However, a quantum mechanical formulation of de Sitter cosmological space-time still lacks in string theory. Recently there was a paper where de Sitter space is considered as a resonance in a scattering process \cite{jonathan}. Classical solutions in Type IIB string theory have been studied by Dasguptha et.al \cite{das}. Another work on  classical de Sitter solutions in string  theory has been published by Danielsson et.al. \cite{dan}. There are several other reasons   to study de Sitter spaces. In high energy physics,  it has been shown that there is a holographic duality relating quantum gravity on the de Sitter space to conformal field theory on a sphere of one dimension lower \cite{witten} \cite{stro}; this is  known as dS/CFT correspondence.

Quasi normal modes (QNM) of black holes is an important aspect of the evolution of perturbations of black hole space-times. There have been plethora of works focusing on computing QNM frequencies for fields of various spins in the background geometry including spin 1 electromagnetic perturbations. One of the best reviews on QNM frequencies written is the one  by Konoplya and Zhidenko \cite{kono1}. In this work, we are interested in studying electromagnetic perturbations by the Maxwell's field in the background of a de Sitter black hole in massive gravity. Since the historical observation of gravitational waves by Advanced LIGO in 2015 \cite{ligo}, the need to understand and calculate QNM frequencies has become even more important since QNM are a part of the evolution of black hole mergers. On the other hand, the current level of observations of gravitational waves does not exclude alternative theories of  gravity as discussed in \cite{yunes} \cite{kono9}. Hence there is motivation to study alternative gravity theories such as massive gravity with a massive graviton.

It is true that when two black holes merge in the vacuum that there will not be any electromagnetic counterparts; however, there is the possibility that  such collisions would be followed by gamma-ray bursts from a jet originated in the accretion flow around the remnants of the black hole mergers \cite{loeb} \cite{bran}. This idea have resulted in range of new models that may endow black hole mergers with electromagnetic observations. For example, electromagnetic luminosity of coalescence of charged black hole binaries were studied by Liebling and Palenzuela \cite{steve}. In another study, the coupling between gravitational waves and electromagnetic waves were studied by Cabral and Lobo \cite{lobo}. Sotani et.al. \cite{sota1}\cite{sota2} studied electromagnetic signature driven by gravitational perturbations in black holes and neutron stars. Hence electromagnetic perturbations of black hole space-times will take an important place in perturbation studies of black holes and also computing QNM frequencies. There are few works we would like to mention  along those lines: Electromagnetic QNMs of D-dimensional black holes were done in \cite{lopez}. Electromagnetic and Dirac perturbations of a Ho$\check{r}$ava black hole evolution was   presented in \cite{var}. Electromagnetic perturbations with Weyl corrections were done by Chen and Jing \cite{chen} and of black hole surrounded by quintessence was done by \cite{zhang}.

Massive gravity theories are extensions of general relativity with a graviton which has a mass. There are many theories of massive gravity in the literature; a nice review is given by de Rham \cite{rham}. One can divide all the theories of massive gravity into two broad categories; the Lorentz invariant theories \cite{ivan} and the Lorentz breaking theories. In this paper we will focus on a black hole arising from a Lorentz breaking theory, which will be described  in detail in the next section.

The paper is organized as follows: in section 2 an introduction to the de Sitter black hole in massive gravity is given. In section 3 the electromagnetic  field perturbation is introduced. In section 4, the WKB approach to compute QNM is introduced.  In section 5, P$\ddot{o}$schl-Teller method to compute QNM frequencies is presented.  In section 6, absorption cross sections are calculated. In section 7 the conclusion is  given.


\section{ Introduction to massive black hole in de Sitter space}

The Lorentz breaking massive gravity theory considered in this paper is obtained from the following action,

\be \label{action}
S = \int d^4 x \sqrt{-g } \left[ - M_{Pl}^2 \mathcal{R}  + L_m + \alpha^4 \mathcal{F}( X, W^{ij}) \right]
\ee
The first two terms are the standard general relativity terms; the curvature term and the Lagrangian of ordinary matter minimally coupled to gravity. The third term, $\Lambda^4 \mathcal{F}$ is a functional of four scalar fields, $\phi^0, \phi^i$. These scalar fields will break the Lorentz symmetry when they acquire a space-time depending vacuum expectation value. These scalar fields are known as Goldstone fields. The functions $X$ and $W^{ij}$ are given in tersm of the scalar fields as below:
\be
X = \frac{\partial^{\mu} \phi^0 \partial_{\mu} \phi^0 }{ \alpha^4}
\ee
\be
W^{i j} = \frac{\partial^{\mu} \phi^i \partial_{\mu} \phi^j}{ \alpha^4} - \frac{\partial^{\mu} \phi^i \partial_{\mu} \phi^0  \partial^{\nu} \phi^j \partial_{\nu} \phi^0 }{ \alpha^8 X}
\ee
The action in eq$\refb{action}$ can be treated as a low-energy effective theory below the ultraviolet cutoff below the scale given by $\alpha$. Here the scale $\alpha$ has the dimensions of mass and it is of the order of $\sqrt{m M_{Pl}}$. The mass of the graviton is $m$ and $M_{Pl}$ is the Planks mass \cite{dubo}\cite{ruba2}\cite{luty}\cite{ruba}\cite{pilo1}\cite{tinya}\cite{pilo}.

The black hole solution was derived for the above theory in \cite{tinya}. We will omit the details of the derivation and present the geometry here. The metric corresponding to the black hole is given by,
\begin{equation} \label{metric}
ds^2 = - f(r) dt^2 + \frac{ dr^2}{ f(r)} + r^2 ( d \Omega_2^2)
\end{equation}
where,
\begin{equation} \label{fr}
f(r) = 1 - \frac{ 2 M} { r} - \gamma \frac{ Q^2}{r^{\beta}} - \frac{ \Lambda r^2}{3}
\end{equation}
The scalar fields $\phi^0, \phi^i$ are given by,
\be
\phi^0 = \kappa^2 \left( t + h(r) \right); \hspace{1 cm} \phi^i =   \kappa^2 x^i
\ee
where
\be \label{hr}
h(r) = \pm \int \frac{ dr} { f(r)} \left[ 1 - f(r) \left( \frac{ \gamma Q^2 \lambda(\lambda-1)}{ 12 m^2 b^6}\frac{1}{ r^{\lambda+2}} + 1 \right)^{-1} \right]^{1/2}
\ee
The constant $\beta$ in eq.$\refb{fr}$ is an integration constant and is positive. When $\beta <1$, the ADM mass of the black hole solution diverges. For $\beta \geq 1$, the metric approaches the usual Schwarzschild-de Sitter metric for large distances with finite mass $M$.  Hence we will choose $\beta \geq 1$ for the rest of the paper. The constant $b$ in eq$\refb{hr}$ is related to the cosmological constant as $\Lambda = 2 m^2 ( 1 - b^6)$. In the derivation of the above black hole solution \cite{tinya}, the cosmological constant term $- \Lambda/3 r^2$ was not in the function $f(r)$. There $ b$ was chosen to be 1. However, it is possible to choose $ b \neq 1$ so that the cosmological term exists. In this paper we will choose $ b <1$ so that $ \Lambda >0$ leading to de Sitter geometry.

The constant $\gamma = \pm 1$. When $\gamma =1$ the geometry is very similar to the Schwarzschild-de Sitter black hole.  For $\gamma =-1$, the geometry is very similar to the Reissner-Nordstrom-de Sitter black hole. For the rest of the paper, we will focus on $\gamma =1$ solutions only. 

The Hawking temperature of the massive gravity black hole is given by,
 \be \label{temp}
 T =  \frac{ 1}{ 4 \pi}  \left| \frac{ df(r)}{ dr} \right|_ { r = r_b} = \frac{1}{ 4 \pi} \left(\frac{ 2 M}{ r_b^2} + \frac{ \gamma Q^2 \lambda}{ r_b^{ \lambda + 1}}   - \frac{ 2 \Lambda r_b}{ 3}\right)
 \ee
 Here $r_b$ is the black hole event horizon. Since at the black horizon $ f(r_b) = 0$, the mass of the black hole could be written as,
 \be \label{mass}
 M = \frac{r_b}{2} - \frac{ \gamma Q^2}{ 2 r_b^{( \lambda -1)} }-  \frac{ r_b^3 \Lambda}{6}
 \ee
The temperature is plotted by varying the scalar charge $Q$  in Fig$\refb{temp}$: one can observe that there is a maximum for the temperature.
\begin{figure} [H]
\begin{center}
\includegraphics{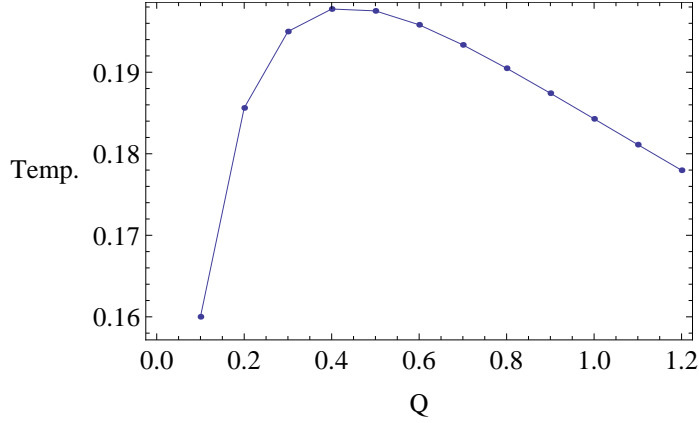}
\caption{The figure shows  $Temperature$ vs $Q$. Here $M = 0.3, \Lambda =0.2$, and  $\beta =6$}.
\label{temp}
 \end{center}
 \end{figure}


\section{ Electromagnetic field  perturbations}

The evolution of a sourceless electromagnetic field in the black hole background geometry is given by the Maxwell's equation,
\be \label{max}
\bigd_{\mu} F^{\mu \nu} = 0 
\ee
Here the electromagnetic field tensor $F_{\mu \nu}$ and the potential $A_{\mu}$ are related by,
\be
F_{\mu \nu} = \bigd_{\mu} A_{\nu} - \bigd_{\nu} A_{\mu}
\ee
Since the background geometry is spherically symmetric, one can deompose the function $A_{\mu}$ in terms of vector spherical harmonics as \cite{ruf}
\be \label{exp}
A_{\mu}(t,r,\theta,\phi) = \sum_{\ell,m}
\left(\left[\begin{array}{c}
0\\
0 \\
\frac{\delta(r,t)}{\sin\theta}\ \frac{\partial Y^{\ell m}}{\partial \phi}\\
-\delta(r,t)\ \sin\theta\ \frac{\partial Y^{\ell m}}{\partial \theta}
\end{array}\right] \right. \nonumber + \left. \left[\begin{array}{c}
\eta(r,t) Y^{\ell m} \\
\kappa(r,t) Y^{\ell m} \\
\xi(r,t)\ \frac{\partial Y^{\ell m}}{\partial \theta} \\
\xi(r,t)\ \frac{\partial Y^{\ell m}}{\partial \phi}
\end{array}\right]
\right) \,\, 
\ee
Here, $Y_{l,m} (\theta, \phi)$ are spherical harmonics, and, $l$ and $m$ are the angular and the azimuthal quantum numbers respectively. The first column in eq.$\refb{exp}$ is the axial component with parity $(-1)^{l+1}$ and the second term is the polar mode with parity $(-1)^l$.  In \cite{molina}, it was demonstrated that both the axial and the polar modes of the electromagnetic perturbations simplifies to an equation,
\be \label{wave}
\frac{ d^2 \Psi(r_*) }{ dr_*^2} + \left( \omega^2 - V_{eff}(r_*)  \right) \Psi(r_*) = 0
\ee
Here  $r_*$ is the tortoise coordinates which is given by,
\be \label{tor}
dr_* = \frac{ dr} { f(r)}
\ee
and  $V_{eff}$ is given by,
\be \label{potem}
V_{eff}(r) = f(r) \left( \frac{ l ( l + 1) } { r^2} \right)
\ee
Here, the function $\Psi(r_*)$ takes different forms for the two modes: for odd parity,
\be
\Psi(r) = \delta(r)
\ee
and for even parity,
\be
\Psi(r) =  -\frac{r^2}{ l ( l+1)} \left( i \omega \kappa(r) +  \frac{ d \eta}{dr} \right) 
\ee
Note that the time dependence for all the above functions take the form $ \Psi(r, t) = \Psi(r) e^{- i \omega t}$. Here $\omega$ is the oscillating frequency of the electromagnetic wave.


\subsection{ Effective potential}

\begin{figure} [H]
\begin{center}
\includegraphics{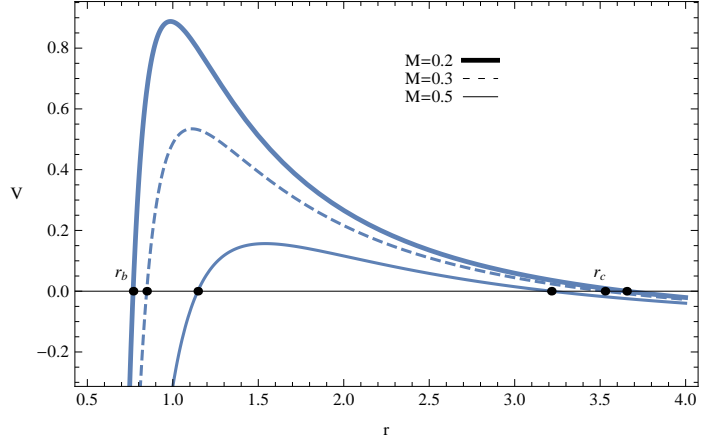}
\caption{The figure shows  $V_{eff}(r)$ vs $r$. Here $Q = 0.3, \Lambda = 0.2, \beta =6$ and $l=1$}
\label{potmass}
 \end{center}
 \end{figure}

\begin{figure} [H]
\begin{center}
\includegraphics{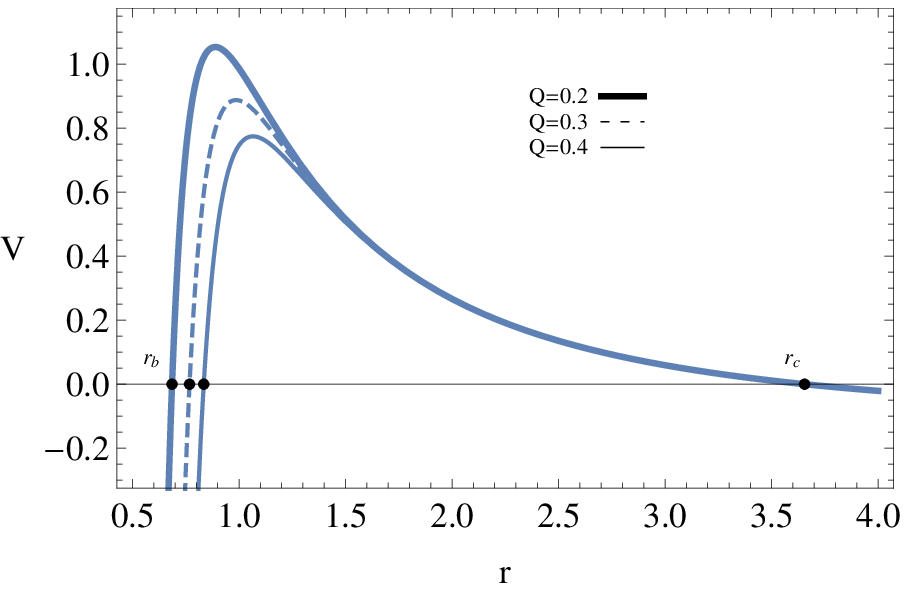}
\caption{The figure shows $V_{eff}(r)$ vs $r$. Here $M = 0.2, \Lambda = 0.2, \beta =6$ and $l=1$}
\label{potcharge}
 \end{center}
 \end{figure}

The effective potential $V_{eff}$ approaches zero at $ r = r_b$ ad $r = r_c$. It has a peak in between the two horizons as demonstrated in Fig.$\refb{potmass}$ and it is positive between the horizons. When $r \ra r_b$, $r_* \ra - \infty$ and when $r \ra r_c$, $r_* \ra \infty$.
Hence the effective potential $V_{eff} \ra 0$ when $ r_* \ra \pm \infty$. In the following figures, we have plotted the $V_{eff}$ as a function of $r$. In this case, it depends on the parameters, $M, Q, \beta, \Lambda$ and $l$. In Fig$\refb{potmass}$, the potential is plotted by varying the mass $M$: when the mass increases, the height of the potential decreases. The potential by varying the charge is plotted in Fig$\refb{potcharge}$: here it is clear that when the charge increases, the height decreases. When the cosmological constant is increased, the height of the potential decreases. However, the position of $r$ where the peak occur is the same for all $\Lambda$. One can understand this simply by observing the potential as,
\be
V_{eff}(r) = l ( l+1) \left( \frac{1}{r^2} - \frac{2 M}{ r^3} - \frac{Q^2}{ r^{\beta + 2}} \right)- \frac{ l ( l+1) \Lambda}{3} 
\ee
Hence, the solution for $\frac{d V_{eff}}{dr} =0$ is independent of $\Lambda$. The height of the peak will vary by $\frac{l(l+1) \Lambda}{3}$ when $\Lambda$ is changed.

\begin{figure} [H]
\begin{center}
\includegraphics{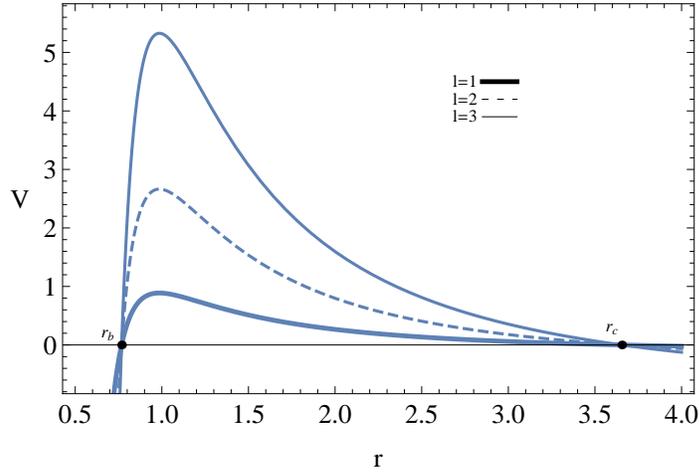}
\caption{The figure shows  $V_{eff}(r)$ vs $r$. Here $M = 0.2, Q = 0.3, \beta =6$ and $\Lambda=0.2$}
\label{potl}
 \end{center}
 \end{figure}

\begin{figure} [H]
\begin{center}
\includegraphics{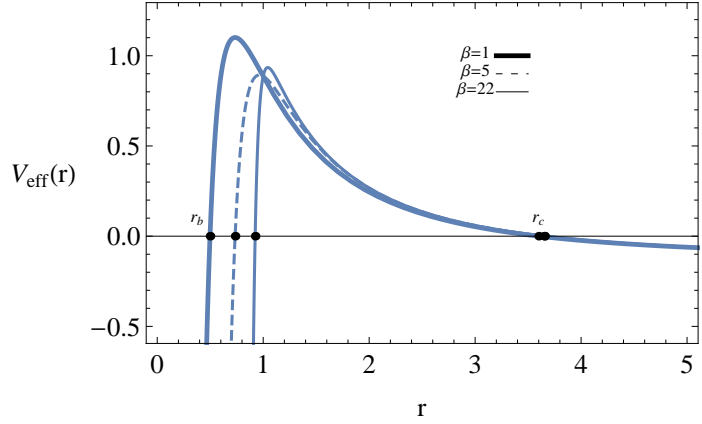}
\caption{The figure shows  $V_{eff}(r)$ vs $r$. Here $M = 0.2, Q = 0.3, l=1$ and $\Lambda=0.2$}
\label{potbeta}
 \end{center}
 \end{figure}

\begin{figure} [H]
\begin{center}
\includegraphics{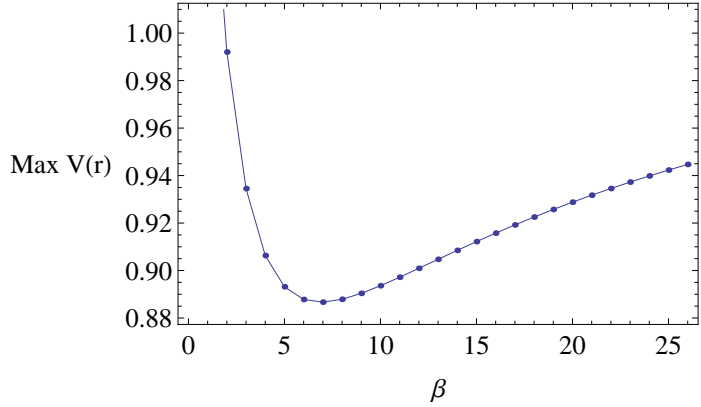}
\caption{The figure shows  $MaxV_{eff}(r)$ vs $\beta$. Here $M = 0.2, Q = 0.3, l=1$ and $\Lambda=0.2$}
\label{potmax}
 \end{center}
 \end{figure}

When $l$ is increased, the height increases as in Fig$\refb{potl}$. The behavior of the potential for varying $\beta$ is quite different from the above. For smaller $\beta$ values, the potential has a higher peak. When the $\beta$ is increased, the potential height decreases, but after a certain value of $\beta$, the height starts to increases again as in Fig$\refb{potbeta}$. This behavior is clarified better in the graph of Fig.$\refb{potmax}$ where the the height of the peak is plotted agianst $\beta$. It is clear that after a certain value of $\beta$, the height increases.

\begin{figure} [H]
\begin{center}
\includegraphics{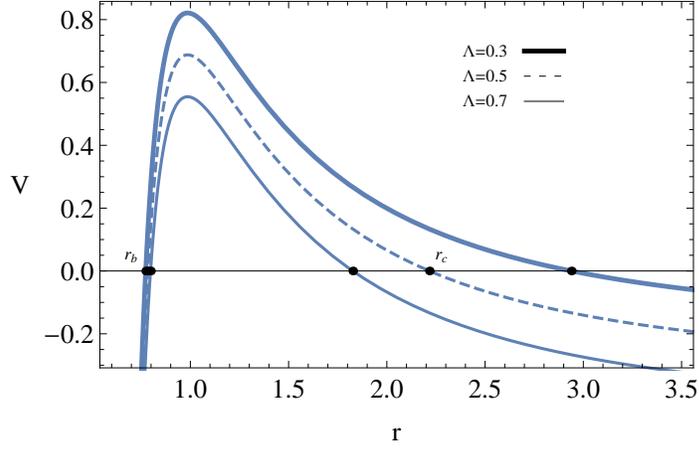}
\caption{The figure shows  $V_{eff}(r)$ vs $r$. Here $M = 0.2, Q = 0.3, \beta =6$ and $l=1$}
\label{potlambda}
 \end{center}
 \end{figure}
 
\section{ Computation of QNM frequencies  of the electromagnetic field perturbations by WKB approach}

QNM for a black hole perturbed by a electromagnetic  field is given by the solutions of the wave equation in eq.$\refb{wave}$ subjected to boundary conditions: $\Psi$ is  purely ingoing  at the even horizon $r_b$ and purely outgoing  at the cosmological horizon $r_c$.  The frequencies corresponding to the QNM are complex and are given by $ \omega = \omega_r + i \omega_i$ where  $\omega_r$ is the oscillating  component  and $\omega_i$ is the damping component of the frequency. QNM for eq.$\refb{wave}$ with the above mentioned boundary conditions can be represented as,
\be
\Psi(r_*) \ra exp( i \omega r_*);  \hspace{1 cm}  r_* \ra - \infty ( r \ra r_b)
\ee
\be
\Psi(r_*) \ra exp( -i \omega r_*);  \hspace{1 cm}  r_* \ra + \infty ( r \ra r_c)
\ee
It is not easy to compute $\omega$ exactly since it is rare for eq$\refb{wave}$ to have exact solutions. We would like to mention here few examples where $\omega$ have been computed exactly for the sake of completeness. For the dilaton black holes in 2+1 dimensions, the exact QNM frequencies have been computed by Fernando \cite{fernando1} \cite{fernando2} \cite{fernando3}.

$\omega$ values for the current work has to be computed using a semi-analytical method, the WKB approach,  to compute QNM frequencies.  This time independent method is very efficient when the potential has a single peak.
WKB approach to find QNM frequencies of black holes was first developed by Schutz, Iyer and Will \cite{will} \cite{will2}. Then it was   extended to sixth order by Konoplya \cite{kono4}. The sixth order WKB approximation is employed to find QNM frequencies in several papers including, \cite{gauss} \cite{fernando4}  \cite{fernando5}. The WKB high order formula for the QNM frequencies are given by,
\be
\omega^2 = - i \sqrt{ - 2 V''(r_{max})} \left( \sum^6_{i=2}  L_i  +   n + \frac{ 1}{2} \right) + V(r_{max})
\ee
Here, $r_{max}$ is  where $V_{eff}(r)$ reach a  maximum and $V''(r)$ is the second derivative of  $V_{eff}(r)$.  Expressions for $L_i$ can be found in \cite{kono4}.  Here $n$ is the mode of the oscillations. We will mainly focus on  the fundamental frequency $n=0$. Note that the WKB approach gives best  accuracy when   $ l >n$. Hence $l$ is chosen to be greater than $n$ in the following computations. All computed $\omega_i$ came out to be negative; the electromagnetic field is stable in the background of the massive gravity black hole. In the subsequent figures, $\omega_r$ and $\omega_i$ is plotted against various parameters in the theory such as $M, Q, \Lambda, l, n$ and $\beta$. Only the magnitude of $\omega_i$ is used in plotting these graphs.

\begin{figure} [H]
\begin{center}
\includegraphics{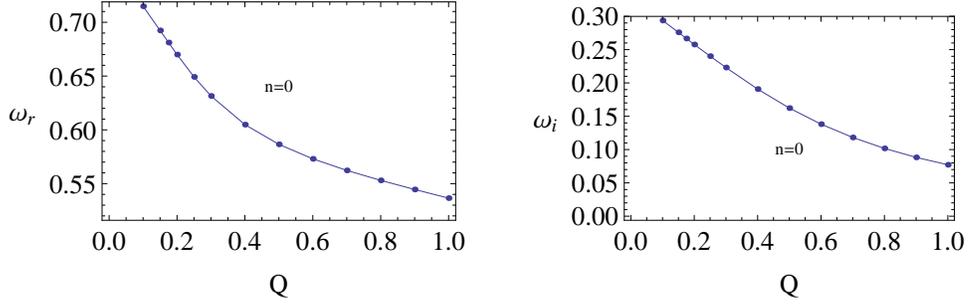}
\caption{The figure shows  $\omega_r$ vs $Q$. Here $M = 0.3, \Lambda = 0.2, l =1, n=0$, and  $\beta =6$}.
\label{omegaq}
 \end{center}
 \end{figure}

First, let us comment on $\omega$ vs $Q$ as plotted in Fig$\refb{omegaq}$. It is clear that $\omega_r$ and $\omega_i$ decreases with $Q$. Hence the electromagnetic field is more stable for smaller $Q$. Hence the massive gravity black hole is more stable than the Schwarzschild-de Sitter black hole.

\begin{figure} [H]
\begin{center}
\includegraphics{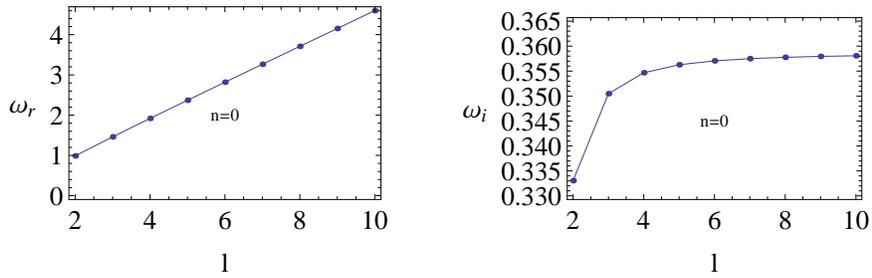}
\caption{The figure shows  $\omega_r$ vs $l$. Here $M = 0.3, \Lambda = 0.2, Q = 0.7$, and  $\beta =6$}.
\label{omegal}
 \end{center}
 \end{figure}

Next, $\omega$ is plotted against $l$ in Fig$\refb{omegal}$. Here, $\omega_r$ has a linear relation with $l$. On the other hand, $\omega_i$ increases and reach a constant value for large $l$. Hence the field decays faster for larger values of $l$.

\begin{figure} [H]
\begin{center}
\includegraphics{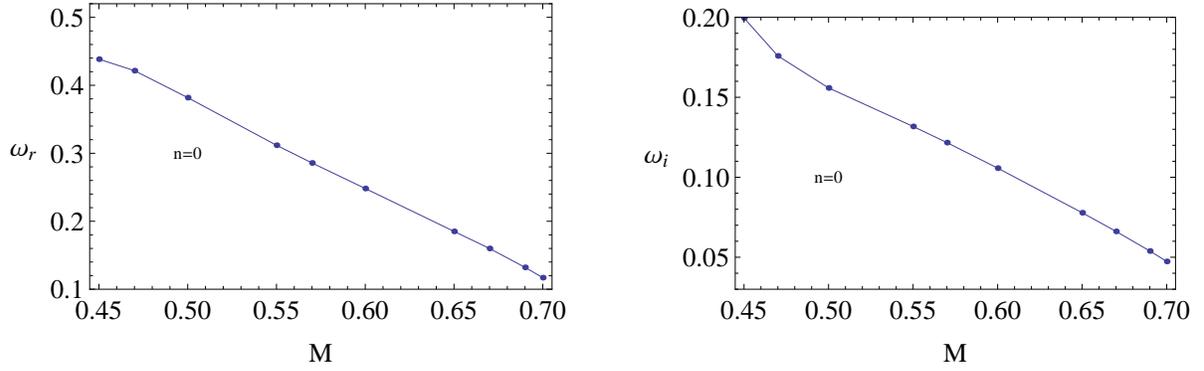}
\caption{The figure shows  $\omega_i$ vs $M$. Here $l= 1, n=0, \Lambda=0.2, Q = 0.7$, and  $\beta =6$}.
\label{omegamass}
 \end{center}
 \end{figure}

When $\omega$ is studied against $M$, as represented in Fig$\refb{omegamass}$, one could observe that both $\omega_r$ and $\omega_i$ decreases. Hence the field is more stable for larger black holes.

\begin{figure} [H]
\begin{center}
\includegraphics{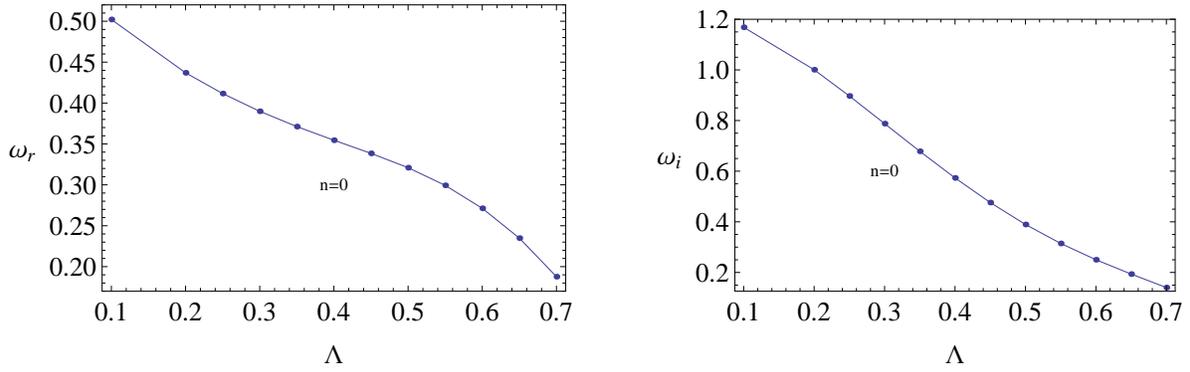}
\caption{The figure shows  $\omega_r$ vs $\Lambda$. Here $l= 1, n=0, M=0.3, Q = 0.7$, and  $\beta =6$}.
\label{omegalambda}
 \end{center}
 \end{figure}

Next, we computed $\omega$ by varying $\Lambda$ as plotted in Fig.$\refb{omegalambda}$. Both $\omega_r$ and $\omega_i$ decreases as $\Lambda$ increases.  Hence, a larger $\Lambda$ leads to a slow decay of the electromagnetic field around the black hole. 

\begin{figure} [H]
\begin{center}
\includegraphics{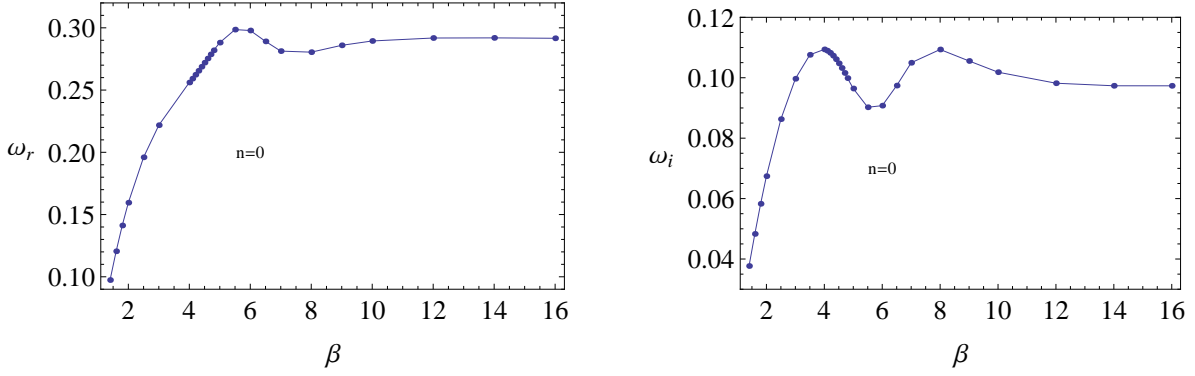}
\caption{The figure shows  $\omega_r$ and $\omega_i$  vs $\beta$. Here $l= 1, n=0,  M= 1, Q = 3, \Lambda =0.001$}.
\label{omegabeta}
 \end{center}
 \end{figure}

The variation of $\omega$ with respect to $\beta$ is plotted in Fig$\refb{omegabeta}$. The frequencies are computed for $n=0$ value. For small values of $\beta$, both $\omega_r$ and $\omega_i$ increases. When $\beta$ gets larger, $\omega_r$ reach a maximum, decreases, and reach a stable value.  $\omega_i$ also increases to a maximum but reach a minimum, and then reach a second maximum before approaching a stable value. This unusual behavior may be due to the fact that the effective potential in Fig$\refb{potbeta}$ has a minimum. The stable value reached for both $\omega_r$ and $\omega_i$ are the value corresponding to the Schwarzschild-de Sitter black hole which is $\omega = 0.29134 - i 0.0974723$.

\begin{figure} [H]
\begin{center}
\includegraphics{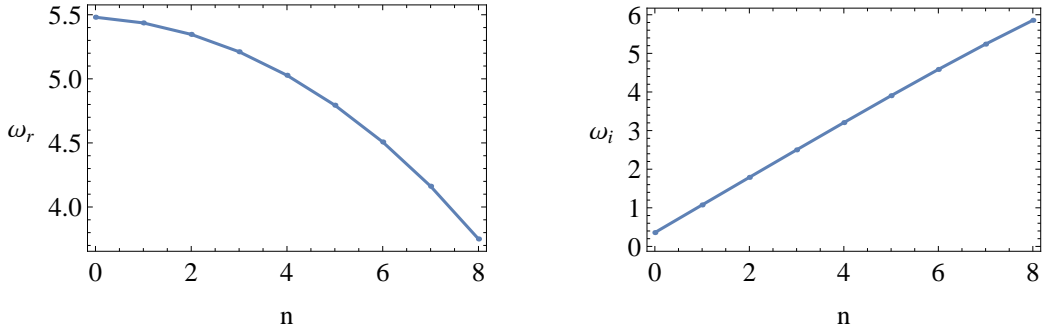}
\caption{The figure shows  $\omega_r$ and $\omega_i$ vs $n$. Here $l= 12, M=0.3, Q = 0.7, \Lambda =0.2, \beta =3$}.
\label{omegan}
 \end{center}
 \end{figure}

The behavior  of the frequencies for higher harmonics is plotted in Fig$\refb{omegan}$. $\omega_r$ decreases for large $n$. Hence the oscillation is less for higher harmonics. $\omega_i$ have a linear relation with $n$. Hence the electromagnetic field decays faster for higher harmonics.

\subsection{ Comparison of frequencies between  the Schwarzschild-de Sitter black hole and the massive gravity black hole }

In this section we present values for frequencies for the Schwarzschild-de Sitter black hole and the massive gravity black hole. The data for the SdS black hole is from the paper by Zhidenko \cite{zhi}.  $\omega_r$ amd $\omega_i$ are calculated for $ l=1$ and $l=2$.  $\omega_r$ is larger for the SdS black hole: the waves oscillates more in the back ground of the SdS black hole. $\omega_i$ is smaller for the massive gravity black hole: the wave decays faster in the SdS black hole background.

\begin{center}
\begin{tabular}{|l|l|l|l|l|r} \hline \hline
 $\Lambda$ & $\omega_r$(SdS BH)  &  $\omega_i$ (SdS BH) & $\omega_r$(Massive-dS BH) 
&  $\omega_i$ (Massive-dS BH) \\ \hline

$0$&0.2482 & - 0.0926 i& 0.2470 & - 0.0926 i\\ \hline
$0.02$&0.2259 & - 0.0842 i& 0.2247 & - 0.0842 i \\ \hline
$0.04$&0.2006 &- 0.0748 i & 0.1992 & - 0.0747 i\\ \hline
$0.06$&0.1709 &- 0.0639 i & 0.1694 & - 0.0636 i \\ \hline
$0.08$&0.1339 &- 0.0502 i & 0.1320 & - 0.0498 i\\ \hline
$0.09$&0.11053 &- 0.04156 i & 0.1083 & - 0.0409 i\\ \hline
$0.10$&0.08035 &- 0.03028 i & 0.07719 & - 0.0292 i\\ \hline
$0.11$&0.02545 &- 0.00962 i & 0.01247 & - 0.00474 i \\ \hline
\hline
\end{tabular}\\
\vspace{0.5 cm}
Table 1: $\omega_r$ and $\omega_r$ for  SdS and Massive dS black holes: l =1, n=0. Here $M =1$.
\end{center}

\begin{center}
\begin{tabular}{|l|l|l|l|l|r} \hline \hline

 $\Lambda$ & $\omega_r$(SdS BH)  &  $\omega_i$ (SdS BH) & $\omega_r$(Massive-dS BH) 
&  $\omega_i$ (Massive-dS BH) \\ \hline

\hline
   $0$&0.45759&- 0.09501 i & 0.45572 & - 0.09504 i\\ \hline
$0.02$&0.41502 &- 0.08615 i & 0.41299 & - 0.08611 i\\ \hline
$0.04$&0.36723 &- 0.07624 i& 0.36496 & - 0.07611 i\\ \hline
$0.06$&0.31182 &- 0.06478 i & 0.30917 & - 0.06452 i\\ \hline
$0.08$&0.24365&- 0.0506 i & 0.24027 & - 0.05020 i\\ \hline
$0.09$&0.20085&- 0.04180 i & 0.19676 & - 0.04114 i\\ \hline
$0.10$&0.14582&- 0.03037 i & 0.14015 & - 0.02933 i\\ \hline
$0.11$&0.04614&- 0.00962 i & 0.02261 & - 0.00473 i\\ \hline
\hline
\end{tabular}\\

\vspace{0.5 cm}
Table 2: $\omega_r$ and $\omega_r$ for  SdS and Massive dS black holes: l =2, n=0. Here $M =1$.
\end{center}


\begin{center}
\begin{tabular}{|l|l|l|l|l|r} \hline \hline

 $\Lambda$ & $\omega_r$(SdS BH)  &  $\omega_i$ (SdS BH) & $\omega_r$(Massive-dS BH) 
&  $\omega_i$ (Massive-dS BH) \\ \hline

\hline
   $0$&0.43653&- 0.29073 i & 0.43445 & - 0.29088 i \\
$0.02$&0.39900&- 0.26202 i & 0.39684 & - 0.26192 i\\
$0.04$&0.35602&- 0.23065 i & 0.35371 & - 0.23025 i\\
$0.06$&0.30498&- 0.19516 i & 0.30235 & - 0.19436  i\\
$0.08$&0.24046&- 0.15223 i & 0.23713 & - 0.15080 i \\
$0.09$&0.19907&- 0.12549 i & 0.19506 & - 0.12348 i\\
$0.10$&0.14515&- 0.09114 i & 0.13954 & - 0.08799 i\\
$0.11$&0.04612&- 0.02886 i & 0.02261 & - 0.01420 i \\
\hline
\end{tabular}\\

\vspace{0.5 cm}
Table 3: $\omega_r$ and $\omega_r$ for  SdS and Massive dS black holes: l =2, n=1. Here $M =1$.
\end{center}


\section{ P$\ddot{o}$schl-Teller approximation for the near-extreme  de Sitter black hole in massive gravity}

For certain parameters of the theory, the black hole can have degenerate horizons with $r_b = r_c$. For such black holes, two conditions has to be satisfied:
\be 
f(r) =0;  \hspace{ 1 cm}  f'(r) =0
\ee
Lets say the mass of the black hole in this case is $M_{cri}$. In this case, the extreme black hole radius $\rho$ is a solution of the equation,
\be
 ( 2 + \beta) \Lambda r^3 - 3 r \beta + 6 M_{cri} ( \beta -1) =0
 \ee
 At $\rho$, 
 \be
 f''(\rho) = \frac{ 6 M_{cri} ( \beta -1) - 2 \rho^3 ( 2 + \beta ) \Lambda}{ 3 \rho}
 \ee
 It is clear that $f''(\rho) <0$ due to the nature of the function $f(r)$ at $r = \rho$. Such extreme black holes in de Sitter geometry are called Nariai black holes \cite{mat2} \cite{fernando6} \cite{fernando7} \cite{fernando8}. The topology near the degenerate horizon is $dS_2 \times S^2$.  $\Lambda_{effective}$ for the $dS_2$ is given by $|f''(\rho)|/2$.
  
When the black holes are near extreme, $f(r)$ can be expanded in a Taylor series as \cite{mat2} as,
\begin{equation}
f(r) \approx  \frac{ f''(\rho)}{2} ( r - r_b) ( r - r_c)
\end{equation}
Hence the tortoise coordinate defined  as $r_* = \int \frac{dr}{f(r)}$ can be integrated to be,
\be \label{newtr}
r_* =   - \frac{1}{\xi} log\left( \frac{ r_c - r}{ r - r_b} \right) 
\ee
Here,
\be
\xi = -\frac{2}{ f''(\rho) ( r_c - r_b)}
\ee
\noindent
From the relation in eq.$\refb{newtr}$, $r$ can be solved to be,
\be \label{rvalue}
r = \frac{ r_b + r_c e^{ \xi r_*}}{ 1 + e^{ \xi r_*}}
\ee
Now, the function $f(r)$ for the near extreme massive gravity-de Sitter black hole can be written as,
\be \label{newfr}
f(r)  =  \frac{ \xi(r_c - r_b)}{ 4( Cosh( \frac{ \xi r_*}{2}))^2} 
\ee
Hence the  the effective potential $V_{eff}$ for the electromagnetic  field can be written as,
\be
V_{eff} =  \frac{l(l+1) f(r)}{r^2} = \frac{ V_0}{ Cosh^2( \frac{\xi r_*}{2})}
\ee
where
\be \label{v0}
V_0 = \frac{ \xi l ( l+1)  ( r_c -r_b)}{ 4 \rho^2} 
\ee
In deriving $V_0$,  we have assumed  $ r  \approx \rho$. 
$\frac{ V_0}{ Cosh^2( \frac{\rho r_*}{2})}$ is the well known   P$\ddot{o}$schl-Teller potential: Ferrari and Mashhoon\cite{ferra3} demonstrated that  $\omega$ can be computed exactly as,
\be
\omega =  \sqrt{ V_0  - \frac{\xi^2}{16}} - i \frac{\xi}{2} ( n + \frac{1}{2})
\ee

When the spherical harmonic index $l$ is large, $V_0 \approx \frac{ \xi l^2}{ 4 \rho^2}$. Hence $\omega_r \approx \frac{ \sqrt{\xi} l} { 2 \rho}$. Therefore $\omega_r$ depends on $l$ linearly  for large $l$; this is clear from 
Fig$\refb{omegal}$. Also, for large $l$, $\omega_i$ is independent of $l$ and is a constant; this behavior is evident  in Fig$\refb{omegal}$. On the other hand, for large $n$, $\omega_i$ becomes large and does depend linearly on $n$.
Here, $\omega_r$ is independent of $n$ which is clear from  Fig.$\refb{omegan}$.


\section{ Absorption cross section at high frequency limit}

In this section we will study the absorption cross section via the null geodesics of the black hole. Suppose the electromagnetic wave is coming from the cosmological horizon where $r_* \ra \infty$. When the wave arrive at the black hole horizon, partial get transmitted and the rest get reflected back to the cosmological horizon. Hence, the solution of the electromagnetic wave closer to the horizon and closer to the cosmological horizon can be represented as,
\be
\Psi(r_*) =  T_l(\omega) e^{ - i \omega r_*} \hspace{1 cm} r_* \ra - \infty( r \ra r_h)
\ee
\be
\Psi(r_*) =   e^{ - i \omega r_*}    +   R_l(\omega) e^{  i \omega r_*}  \hspace{1 cm} r_* \ra   +\infty( r \ra r_c)
\ee
Here, $R_l(\omega)$ and $T_l(\omega)$ are the reflection  and transmission coefficient respectively and are related by, $|R_l(\omega)|^2 + |T_l(\omega)|^2 = 1$.
The absorption cross section can be written as,
\be
\sigma_{abs} = \sum \sigma_l
\ee
where $\sigma_l$ is the partial absorption cross section given by,
\be
\sigma_l = \frac{ \pi ( 2 l + 1)}{ \omega^2} |T_l(\omega)|^2
\ee

\begin{figure} [H]
\begin{center}
\includegraphics{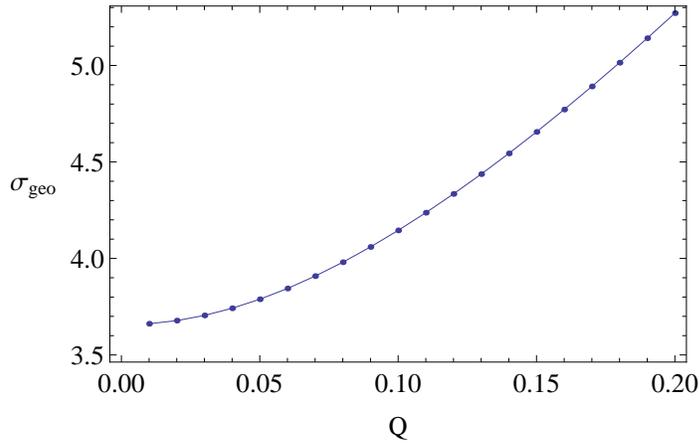}
\caption{The figure shows  $\sigma_{geo}$ vs $Q$. Here $\beta= 3, M=0.2,  \Lambda =0.2$}.
\label{sigmaq}
 \end{center}
 \end{figure}

Since electromagnetic waves represents null geodesics, one can use the classical capture cross section or the geometric cross section of null geodesics to calculate the absorption cross section. Let us briefly introduce the equations of null geodesics as follows. The equations of motion of the null geodesics for a static spherically symmetric black hole  is given by,
\begin{equation} \label{lag}
\mathcal{L}_{null} =  - \frac{1}{2} \left( - f(r) \left( \frac{dt}{d\tau} \right)^2 +  \frac{1}{f(r)}\left( \frac{dr}{d \tau} \right)^2 + r^2 \left(\frac{d \varphi}{d \tau} \right)^2 \right)
\end{equation} 
Here we have considered the motion to be in the plane $\theta = \pi/2$. $\tau$ is an affine parameter. Due to the existence of two Killing vectors, $\partial_{\varphi}$ and $\partial_{t}$, there are two conserved quantities, $L$ and $E$ respectively. They are related to the geometry of the black hole as,
\begin{equation}
r^2 \dot{\varphi} = L; \hspace{1.0cm} f(r) \dot{t} = E
\end{equation}
Then the equation of motion eq$\refb{lag}$ will simplifies to
\begin{equation}
\dot{r}^2 + f(r) \left(  \frac{L^2}{r^2}    \right)  = E^2
\end{equation}
Now, we can represent the effective potential for the null geodesics as, $V_{null} =  \frac{ f(r) L^2}{r^2}$. At the high-frequency limit, the absorption cross section is given by the geometric cross section $\sigma_{geo}$ given by,
\be
\sigma_{geo} = \pi b_c^2
\ee
where
\be
b_c =  \sqrt{\frac{  r_{c}^2}{ f(r_{c})}}
\ee
Here $r_c$ is the radius of the unstable circular orbit of the photons obtained from  $\frac{ dV_{null}}{dr} =0$. $b_c$ is the critical parameter given by $b_c = \frac{L_c}{E_c}$. Both $L_c,E_c$ corresponds to the values for the circular orbit. In Fig$\refb{sigmaq}$, the capture cross section $\sigma_{geo}$ is plotted by varying the scalar charge $Q$. One can observe that  $\sigma_{geo}$ increases with $Q$. Hence the massive gravity black hole absorbs more than the Schwarzschild-de Sitter black hole (the black hole with Q =0).

\begin{figure} [H]
\begin{center}
\includegraphics{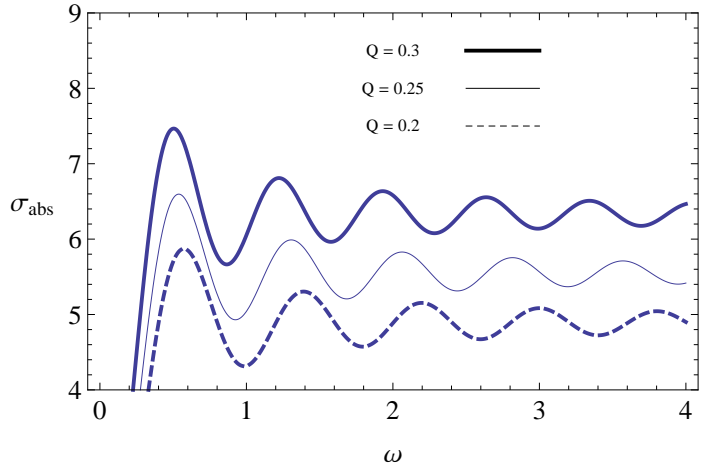}
\caption{The figure shows  $\sigma_{abs}$ vs $\omega$. Here $\beta= 2, M=0.2,  \Lambda =0.2$}.
\label{absq}
 \end{center}
 \end{figure}

\begin{figure} [H]
\begin{center}
\includegraphics{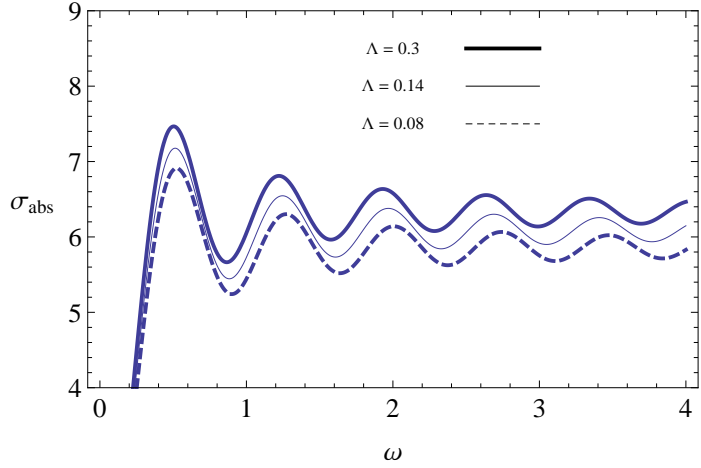}
\caption{The figure shows  $\sigma_{abs}$ vs $\omega$. Here $\beta= 2, M=0.2,  Q =0.3$}.
\label{abslambda}
 \end{center}
 \end{figure}

D$\grave{e}$canini et.al. \cite{deca} demonstrated that the absorption cross section at high-frequency can be improved to be written as,
\be
\sigma_{abs}^{hf} \approx \sigma_{geo} + \sigma_{oscillating}
\ee
Here,
\be
\sigma_{oscillating} = - \frac{ 4 \Gamma_l} { \omega \Omega_l^2} e^{-\frac{\Gamma_l}{\Omega_l} }sin \left( \frac{ 2 \pi \omega}{ \Omega_l} \right)
\ee
In the above expression, $\Gamma_l = \pi \lambda_l$ where $\lambda_l$ is the Lyapunov exponent of the null geodesics given by \cite{cardoso1}\cite{mas1},
\begin{equation} \label{lambdanull}
\lambda_{l} = \sqrt{ \frac{ -V_{null}''(r_{c})}{ 2 \dot{t}(r_{c})^2}} 
= \sqrt{ \frac{-V_{null}''(r_{c}) r_{c}^2 f(r_{c})}{2 L^2} }
\end{equation}
and $\Omega_l$ is the angular velocity of the null geodesics given by,
\begin{equation}
\Omega_a = \frac{ \dot{\phi}(r_{c})}{\dot{t}(r_{c})}  = \sqrt{ \frac{ f(r_{c})}{r_{c}^2 }} 
\end{equation}
In eq.$\refb{lambdanull}$,  $V_{null}''(r) = \frac{ d^2 V_{null}}{dr^2}$.
The above approximation is known as the ${\it sinc \hspace{0.1 cm} approximation}$ in the literature \cite{cris}. In Fig$\refb{absq}$  $\sigma_{abs}^{hf}$  is plotted by varying $Q$. For higher $Q$ the absorption cross section is high. In Fig$\refb{abslambda}$, $\sigma_{abs}^{hf}$ is plotted by varying $\Lambda$. For larger $\Lambda$ the absorption is high.

\subsection{ Comparison of $\sigma_{abs}$ and $\sigma_{geo}$  with the Schwarzschild-de Sitter black hole }

\begin{figure} [H]
\begin{center}
\includegraphics{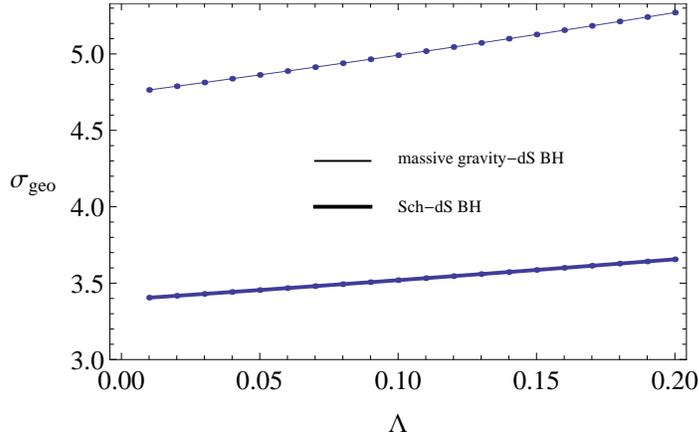}
\caption{The figure shows  $\sigma_{geo}$ vs $\Lambda$ for the massive gravity black hole and the Schwarzschild-de Sitter black hole. Here $\beta= 3, M=0.2,  Q =0.2$}.
\label{sigmalambda}
 \end{center}
 \end{figure}

In Fig$\refb{sigmalambda}$,  $\sigma_{geo}$ is plotted by varying $\Lambda$ for Schwarzschild-de Sitter black hole and the massive gravity black hole. First, when $\Lambda$ increases, $\sigma_{geo}$ increases for both black holes. Hence a large cosmological constant favors a larger absorption cross section. Second, for all values of $\Lambda$,  $\sigma_{geo}$ is smaller for the Schwarzschild-de Sitter black hole.

\begin{figure} [H]
\begin{center}
\includegraphics{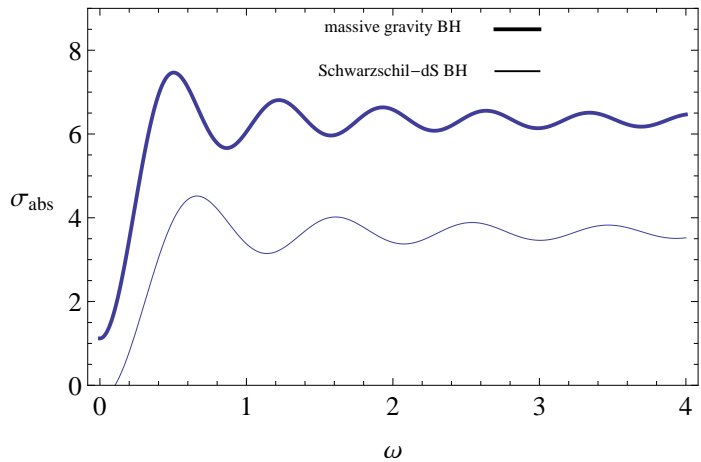}
\caption{The figure shows  $\sigma_{abs}$ vs $\omega$ for both the massive gravity black hole and the Schwarzschild-de Sitter black hole. Here $\beta= 2, M=0.2,  \Lambda =0.2$}.
\label{absboth}
 \end{center}
 \end{figure}

\begin{figure} [H]
\begin{center}
\includegraphics{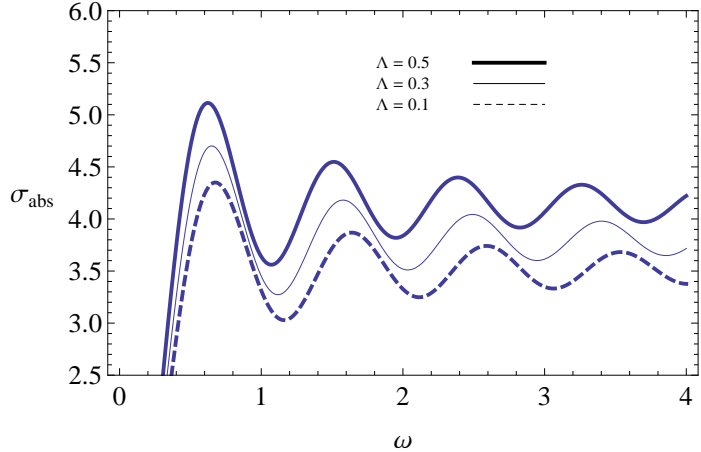}
\caption{The figure shows  $\sigma_{abs}$ vs $\omega$ for the Schwarzschild-de Sitter black hole. Here $M=0.2$}.
\label{schabs}
 \end{center}
 \end{figure}

In  Fig$\refb{absboth}$, $\sigma_{abs}^{hf}$  is plotted for both the Schwarzschild-de Sitter  black hole and the massive gravity black hole. The massive gravity black hole has a higher absorption cross section. In Fig$\refb{schabs}$, $\sigma_{abs}$ at high frequency is plotted for the Schwarzschild-de Sitter black hole by varying $\Lambda$. When $\Lambda$ increases, the absorption is higher similar to the massive gravity black hole. There are only few works that has focused on the absorption (or emission) of fields from the Schwarzschild-de Sitter black hole in the literature. In an interesting work by Kanti et. al. \cite{kanti} absorption cross sections were calculated for  Schwarzschild black holes embedded in $D$ dimensional de Sitter universes. The paper focused on scalar field emission and absorption. For a scalar field, the effective potential is $V_{scalar} =   \frac{ f(r) l ( l + 1) } { r^2}  + \frac{ f(r) f'(r)} {r}$. However, in the high energy regime, the potential for the scalar field and the electromagnetic field is very similar. Hence one can compare the results of the paper by Kanti et.al.\cite{kanti} to the results in this paper effectively. The authors in \cite{kanti} calculated the $\sigma_{geo}$ for $( 4 + n)$ dimensional Schwarzschild-de Sitter black holes where $n$ is the extra dimension. They observed that $\sigma_{geo}$ depends on both $n$ and $\Lambda$. When $n$ is fixed and $\Lambda$ is varied, $\sigma_{geo}$ increases. This is in fact similar to what is observed for the massive gravity black hole  and for the Schwarzschild-de Sitter black hole in $(3 + 1)$ dimensions using the sinc approximation in this paper.

\section{ Conclusion}

Our main goal in this paper has been to study QNM  of a massive gravity black hole in de Sitter universe under electromagnetic  perturbations.  The massive gravity black hole shows similar geometry to Schwarzschild-de Sitter and Reissner-Nordstrom-de Sitter black hole for two values of the parameters in the theory. It can have two or three horizons depending on the values of the theory.

To calculate the QNM frequencies, we have employed sixth order WKB approximation. The parameters of the theory, the mass $M$, scalar charge $Q$, cosmological constant $\Lambda$,  the spherical index $l$, the value $\beta$ and the mode number $n$ were changed to see how QNM frequencies depend on them.

When $Q$ is  increased, $\omega_r, \omega_i$ decreases. Hence, the field decays faster for low $Q$. When studied the behavior of $\omega$ with respect to $l$, the spherical harmonic index is similar to the behavior of other black holes: $\omega_r$  increases linearly with $l$ and $\omega_i$ increases to a stable value for large $l$.  When the  mass $M$ and $\Lambda$ is increased, $\omega_r$ and $\omega_r$ decreases. Hence the field is stable for smaller values of $M$ and $\Lambda$. From the graph for   $\omega$ vs $n$,  one can observe that  $\omega_r$ decreases with $n$ and  $\omega_i$  increases linearly with $n$.

To understand the numerical results better, we have demonstrated that for the near-extremal black hole with $r_b \approx r_c$ the electromagnetic field equation will have the P$\ddot{o}$schl-Teller potential. One can obtain exact results for the frequencies in this method. It is clear that the linear relation for $\omega_r$ for large $l$  in this method verifying the numerical results.  Also, the linear relation between $n$ and $\omega_i$ is also obvious.

We have used the unstable null geodesics approach to compute the absorption cross section at high frequencies.

As future work, it may be interesting to calculate absorption cross sections for all values of $\omega$ using a method such as Runge-Kutta method. Furthermore, it would be interesting to study QNM frequencies for massive vector fields in the massive gravity in de Sitter background.

\vspace{0.5 cm}


{\bf Acknowledgments:}  The author wish to thank R. A. Konoplya for providing the  {\it Mathematica}  file for  the WKB approximation.



\begin{thebibliography}{99}


\bibitem{perl} S. Perlmutter et. al., {\it Measurements of $\Omega$ and $\Lambda$ from 42 high-redshift supernovae}, Astrophys. J. {\bf 517} 565 (1999)

\bibitem{riess} A.G. Riess  et. al., {\it Observational evidence from supernovae for an accelerating universe and a cosmological constant},  Astron. J. {\bf 116} 1009 (1998);
{\it BVRI Light Curves for 22 Type Ia Supernovae}, Astron. J. {\bf 117} 707(1999)

\bibitem{sper} D. N. Spergel et.al. (WMAP Collaboration), {\it Wilkinson Microwave Anisotropy Probe (WMAP) three year results: implications for cosmology}, Astrophys. J. Suppl. {\bf 170} 377 (2007)

\bibitem{teg} M. Tegmark et.al. (SDSS Collaboration) {\it Cosmological parameters from SDSS and WMAP} , Phys. Rev. {\bf D 69} 103501 (2004)

\bibitem{sel} U. Seljak et.al., {\it Cosmological parameter analysis including SDSS Ly$\alpha$ forest and galaxy bias: constraints on the primordial spectrum of fluctuations, neutrino mass, and dark energy}, Phys. Rev. {\bf D 71} 103515 (2005)

\bib{jonathan} J. Maltz, \& L. Susskind, {\it De Sitter as a resonance}, arXiv:1611.00360

\bib{das} K. Dasguptha, R. Gwyn, E. McDonough, M. Mia, \& R. Tatar, {\it de Sitter vacua in Type II B string theory: classical solutions and quantum corrections}, JHEP 054:1407 (2014)

\bib{dan} U. H. Danielsson, S. S. Haque, G. Shiu, \& T. van Riet, {\it Towards classical de Sitter solutions in string theory}, JHEP 0909:114 (2009)

\bib{witten} E. Witten, {\it Quantum gravity in de Sitter space}: hep-th/0106109

\bib{stro} A. Strominger, {\it The dS/CFT correspondence},  hep-th/0106113




\bibitem{kono1}  R. A. Konoplya and A. Zhidenko, {\it Quasinormal modes of black holes: from astrophysics to string theory}, Rev. Mod. Phys. {\bf 83} 793 (2011) 



\bib{ligo} B. P. Abbot et.al.[LIGO Scientific and Virgo Collaboration], {\it Observation of Gravitational Waves from a Binary Black Hole Merger}, Phys. Rev. Lett. {\bf 116}  061102 (2016)


\bib{yunes}  N.  Yunes, K.  Yagi, and F.  Pretorius { \it Theoretical physics implications of the binary black-hole mergers GW150914 and GW151226}, Phys. Rev. {\bf D 94} 084002 (2016)

\bib{kono9} {\it Detection of gravitational waves from black holes: Is there a window for alternative theories?}, Phys. Lett. {\bf B 756} 350, (2016)

\bib{loeb} A. Loeb, {\it Electromagnetic counterparts to black hole mergers detected by LIGO}, 


\bib{bran} V. Branchina, \& M. De Domenico, {\it Simultaneous observation of gravitational and electromagnetic waves}, arXiv: 1604.08530

\bib{steve} S. L.Liebling, \& C. Palenzuela, {\it Electromagnetic luminosity of the coalescence of charged black hole binaries}, Phys. Rev. {\bf D 94} 064046 (2016)

\bib{lobo} F. Cabral, \& F. S. N. Lobo, {\it Gravitational waves and electrodynamics: new perspectives}, arXiv:1603.08157

\bib{sota1} H. Sontani, K. D. Kokkotas, P. Laguna, \& C. F. Sopuerta, {\it Electromagnetic waves for neutron stars and black holes driven by polar gravitational perturbations }, Gen. Rel. Grav. {\bf 46} 1675 (2014)

\bib{sota2} H. Sontani, K. D. Kokkotas, P. Laguna, \& C. F. Sopuerta, {\it Gravitationally driven electromagnetic perturbations of neutron stars and black holes}, Phys. Rev. {\bf D 87} 084018 (2013)

\bib{lopez} A. L$\acute{o}$pez-Ortega, {\it Electromagnetic quasinormal modes of D-dimensional black holes}, Gen. Rel. Grav. {\bf 40} 1379 (2008)

\bib{var} N. Varghese, \& V. C. Kuriakose, {\it Evolution of electromagnetic and Dirac perturbations around a black hole in Ho$\check{r}$ava gravity}, Mod. Phys. Lett. {\bf A 26} 1645 (2011)

\bib{chen} S. Chen, \& J. Jing, {\it Dynamical evolution of the electromagnetic perturbation with Weyl corrections}, Phys. Rev. {\bf D 88} 064058 (2013)

\bib{zhang} Y. Zhang, Y. Gui, \& F. Yu, {\it Quasinormal modes of a Schwarzschild black hole surrounded by free static spherically symmetric quintessence}, Gen. Rel. Grav. {\bf 39} 1003 (2007)




\bib{rham} C. de Rham, {\it Massive gravity}, arXiv:1401.4173

\bib{ivan} H. Kodama and I. Arruut, {\it Stability of the Schwarzschild-de Sitter black hole in the DRGT massive gravity}, Prog. Theor. Exp. Phys. 023E02 (2014)

\bib{dubo} S. L. Dubovsky, {\it Phases of massive gravity}, JHEP {\bf 0410} 076 (2004)

\bib{ruba2} V. A. Rubakov \&  P. G. Tinyakov, {\it Infrared-modified gravities and massive gravitons}, Phys. Usp. {\bf 51} 759  (2008)

\bib{luty} N. Arkani-Hamed, H. Cheng, M.A. Luty \&  S. Mukohyama, {\it Ghost condensation and a consistent infrared modification of gravity}, JHEP {\bf 0405} 074 (2004)

\bib{ruba} V. Rubakov, hep-th/0407104.

\bib{pilo1} D.Blas, D. Comelli, F. Nesti, \&  L. Pilo,  {\it Lorentz Breaking Massive Gravity in Curved Space}, Phys. Rev. {\bf D 80} 044025 (2009)

\bib{tinya}  M. V. Bebronne \& P. G. Tinyakov, {\it Black hole solutions in massive gravity}, JHEP 0904:100, 2009; Erratum-ibid.1106:018, (2011)

\bib{pilo} D. Comelli, F. Nesti \& L. Pilo, {\it Stars and (Furry) black holes in Lorentz breaking massive gravity}, Phys. Rev. {\bf D 83} 084042 (2011)



\bib{ruf} R. Ruffini, J. Tiomno, \& C. Vishveshwara, {\it } Lett. Nuovo Cim. {\bf 3S2} 211 (1972)

\bib{molina} C. Molina, A.B. Pavan, \& T. E. M. Torrejon, {\it Electromagnetic perturbations in new brane world scenarios}, Phys. Rev. {\bf D 93} 124068 (2016)



\bibitem{fernando1} S. Fernando, {\it Spinning dilaton black holes in 2+1 dimensions: quasinormal modes and the area spectrum},  Phys. Rev. {\bf D 79} 124026 (2009)

\bibitem{fernando2} S. Fernando, {\it Quasinormal modes of charged scalars around dilaton black Holes in 2+1 dimensions: exact frequencies},  Phys. Rev. {\bf D 77}  124005 (2008)

\bibitem{fernando3} S. Fernando, {\it Quasinormal modes of charged dilaton black holes in 2+1 dimensions},  Gen. Rel. Grav. {\bf 36} 71 (2004)


\bibitem{will} S. Iyer \& C.M. Will, {\it Black-hole normal modes: A WKB approach. I. Foundations and application of a higher-order WKB analysis of potential-barrier scattering},  Phys. Rev. {\bf D 35}  3621(1987)


\bibitem{will2} B. F. Schutz \& C. M. Will, {\it  Black hole normal modes; A semi-analytic approach},  Astrophys. Jour. {\bf 291} L33 (1985)

\bibitem{kono4} R. A. Konoplya,  {\it Quasinormal behavior of the D-dimensional Schwarzschild black hole and higher order WKB approach}, Phys. Rev. {\bf D 68}  024018 (2003)

\bib{gauss} R. Konoplya, {\it Quasinormal modes of the charged black hole in Gauss-Bonnet gravity},  Phys. Rev. {\bf D 71} 024038 (2005)

\bib{fernando4} S. Fernando \& T. Clark, {\it Black holes in massive gravity: quasinormal modes of scalar perturbations}, Gen. Rel. Grav. {\bf 46 }   1834 (2014)

\bib{fernando5} S. Fernando, {\it Regular black holes in de Sitter universe: scalar field perturbations and quasinormal modes}, Int. Jour. Mod. Phys. {\bf D 24} 1550104 (2015)

\bib{zhi}  A. Zhidenko, {\it Quasi-normal modes of Schwarzschild-de Sitter black holes}, Class. Quant. Grav. {\bf 21} 273 (2004)



\bib{mat2} J. Matyjasek, P. Saurski \& D. Tryniecki, {\it Inside the degenerate horizons of regular black holes}, Phys. Rev. {\bf D 87} 124025 (2013)

\bib{fernando6} S. Fernando, {\it Cold, ultracold and Nariai black holes with quintessence}, Gen. Rel. Grav. {\bf 45} 2053 (2013)

\bibitem{fernando7}  S. Fernando, {\it Nariai black holes with quintessence}, Mod.  Phys. Lett. {A 28} 13550189 (2013)

\bib{fernando8} S. Fernando, {\it Born-Infeld-de Sitter gravity: cold, ultracold and Nariai black holes}, Int. Jour. Mod. Phys. {D 22} 1350080 (2013)

\bibitem{ferra3} V. Ferrari \& B. Mashhoon, {\it New approach to the quasinormal modes of a black hole},  Phys. Rev. {\bf D 30} 295 ( 1984)


\bib{deca} Y. D$\grave{e}$canini, G. Esposito-Far$\grave{e}$se, \& A. Folacci, {\it Universality of high-energy absorption cross sections for black holes}, Phys. Rev. {\bf D 83} 044032 (2011)

\bibitem{cardoso1}   V. Cardoso, A.S. Miranda, E. Berti, H. Witeck \& V.T. Zanchin, {\it Geodesic stability, Lyapunov exponents and quasinormal modes},  Phys.Rev. {\bf D 79} ( 2009) 064016

\bibitem{mas1} B. Mashhoon, {\it  Stability of charged rotating black holes in the eikonal approximation}, Phys. Rev. {\bf D 31}  (1985) 290

\bib{cris} C. F.  Macedo, \& L. C. B. Crispino, {\it Absorption of planar massless scalar waves by Bardeen regular black hole}, Phys. Rev. {\bf D 90} 064001 (2014)

\bib{kanti}  P. Kanti, J. Grain, \& A. Barrau, {\it Bulk and brane decay of a (4 + n)-dimensional Schwarzschild-de Sitter black hole: scalar radiation}, Phys. Rev. {\bf D 71} 104002 (2005)



\end{thebibliography}
\end{document}